\documentclass{ws-ijmpa}
\newcommand{\be}{\begin{equation}}
\newcommand{\ee}{\end{equation}}
\newcommand{\bea}{\begin{eqnarray}}
\newcommand{\eea}{\end{eqnarray}}

\begin{document}

\markboth{K. A. Milton}
{Recent Developments in Quantum Vacuum Energy for Confined Fields}

%
\catchline{}{}{}{}{}
%

\title{Recent Developments in Quantum Vacuum Energy for Confined Fields
}

\author{\footnotesize K. A. MILTON}

\address{Oklahoma Center for High Energy Physics,
University of Oklahoma, Norman, OK 73072 USA}

\maketitle


\begin{abstract}
Quantum vacuum energy entered hadronic physics through the zero-point
energy parameter introduced into the bag model.  Estimates of
this parameter led to apparent discordance with phenomenological fits.
More serious were divergences which were omitted in an {\it ad hoc\/} 
manner.  New developments in understanding Casimir self-stresses, and the
nature of surface divergences, promise to render the situation clearcut.

\end{abstract}

\section{Introduction}	
In the bag model,\cite{bm} the normal vacuum
is a perfect color magnetic conductor,
that is, the color magnetic permeability $\mu$
 is infinite, while the  vacuum
in the interior of the bag is characterized by $\mu=1$. 
This implies that
the color electric and magnetic fields are confined to the interior of the
bag, and that they satisfy the following boundary conditions on its surface
$S$:
${\bf n\cdot E}|_S=0$, ${\bf n\times B}|_S=0$,
where $\bf n$ is a unit normal to $S$.  Now, even in an ``empty'' bag
(i.e., one containing no quarks) there will be nonzero fields present because
of quantum fluctuations.   
This gives rise to a zero-point or Casimir energy.

The bag-model Lagrangian is
\be
{\cal L}_{\rm bag}=({\cal L}_{\rm D}+{\cal L}_{\rm YM}-B)\eta(\overline
\psi\psi),
\label{bm1}
\ee
where ${\cal L}_{\rm D, YM}$ are the Dirac and Yang-Mills Lagrangians,
respectively,
$B$ is the bag constant, and $\eta$ is the unit step function.
 Variation of 
$\mathcal{L}_{\rm bag}$
 leads to the following linear and quadratic boundary conditions
($n^\mu$ is the outwardly directed normal)
$-i\gamma n\psi=\psi$
and $n_\alpha G^{\alpha\beta}=0$, and
$ -\frac14 G^2-\frac12(n\partial)\overline\psi\psi=B$,
 in terms of its quark contribution.

The MIT bag model consists in solving the Dirac equation subject to the
boundary conditions,
which in the simplest realization refer to a
spherical cavity approximation. The interactions, through the gauge
fields, are treated perturbatively in the approximation that 
the strong coupling is regarded as small, $\alpha_s\ll 1$.
The radial wavefunctions are then expressed in terms of spherical Bessel
functions. 
The results of fits, shown in Table 1 are essentially
identical with those found by DeGrand et al.,\cite{bm} Model E.
\newcommand{\ul}{\underline}
\begin{table}[h]
\label{tab1}
\tbl{Fits to hadron masses for various bag models.}
{\begin{tabular}{@{}ccccc@{}} \toprule
Parameters&Experimental&Model&Model&Model\\
and Masses&Values&A&
D&E\\
\hline
$Z$&&$-1.0$&$1.0$&$1.8327$\\
$B^{1/4}$(GeV)&&$0.19526$&$0.16515$&$0.14535$\\
$F$ (GeV$^2$)&&$-0.21923$&$-0.064729$&$0$\\
$\alpha_s$&&$2.16066$&$2.17932$&$2.19897$\\
$m_s$ (GeV)&&$0.2807$&$0.2802$&$0.2797$\\
$R_N$ (GeV$^{-1}$)&&$5.06$&$5.03$&$5.01$\\
$\delta m$ (GeV)&&$0.1325$&$0.1343$&$0.1408$\\
$m_\pi$&$0.138$&$0.3610$&$0.3229$&$0.2833$\\
$m_\eta$&$0.549$&$0.645$&$0.6105$&$0.5583$\\
$m_{\eta^\prime}$&$0.958$&$0.5031$&$0.4702$&$0.4209$\\
$m_K$&$0.4957$&$0.5827$&$0.5489$&$0.4982$\\
$m_N$&$\ul{0.9389}$&$\ul{0.9389}$&$\ul{0.9389}$
&$\ul{0.9389}$\\
$m_\Lambda$&$1.1156$&$1.1022$&$1.1017$&$1.1012$\\
$m_\Sigma$&$1.1931$&$1.1426$&$1.1426$&$1.1422$\\
$m_\Xi$&$1.13181$&$1.2885$&$1.2876$&$1.1867$\\
$m_\rho$&$0.769$&$0.7826$&$0.7826$&$0.7826$\\
$m_\omega$&$\ul{0.7826}$&$\ul{0.7826}$&
$\ul{0.7826}$&$\ul{0.7826}$\\
$m_\phi$&$1.0195$&$1.0733$&$1.0711$&$1.0688$\\
$m_{K^*}$&$0.8921$&$0.9251$&$0.9246$&$0.9229$\\
$m_\Delta$&$\ul{1.232}$&$\ul{1.232}$&
$\ul{1.232}$&$\ul{1.232}$\\
$m_{\Sigma^*}$&$1.3839$&$1.3760$&$1.3761$&$1.3762$\\
$m_{\Xi^*}$&$1.5334$&$1.5229$&$1.5229$&$1.5230$\\
$m_\Omega$&$\ul{1.6725}$&$\ul{1.6725}$&
$\ul{1.6725}$&$\ul{1.6725}$\\
\hline\end{tabular}}\end{table}
The following comments can be offered concerning these fits.
\begin{itemlist}
\item The $Z=-1$ (repulsive) estimate (my best guess in 1983\cite{tf}),
model A, agrees about
as well with the data as does the optimal bag model E.

\item All the models agree more closely with each other than with
the data.

\item Chiral bag models can be used to improve the fits with the pseudoscalars.

\item For models A and D a constant force term is included: (the bag radius
is $a$)
\be  H=BV+\sigma A+Fa+\frac{Z}a+H_{q,g}.
\ee

\item Some details of how $Z$ is calculated will follow.
\end{itemlist}

\subsection{Gluon and quark condensates}

\begin{figure}
\centerline{
\psfig{file=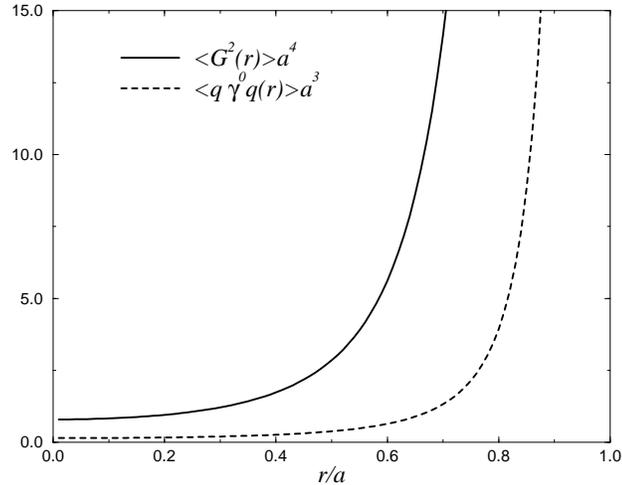,width=3in,angle=270}}
\vspace*{8pt}
\caption{Magnitude of the quark and gluon condensates as functions of the
distance from the center of the bag.  The values shown are for a single
color, and, for the quarks, for a single flavor and helicity state.}
\label{fig1}
\end{figure}

Due to zero-point fluctuations, confinement will result in a nonzero
expectation value for the squared field strength, ($\delta\to0$ is a 
time-splitting regulator)
\be 
\langle G^2(r)\rangle ={1\over4\pi^2 a}{1\over r^2}{d\over dr}\sum_{l=1}
^\infty (2l+1)2\int_0^\infty dx\,e^{ix\delta}{s_l(xr/a)s_l'(xr/a)\over
s_l(x)s_l'(x)},
\ee
Similarly, we obtain the following expression for the quark condensate
\be 
\langle \bar qq(r)\rangle =-{1\over4\pi^2 ar^2}\sum_{l=0}^\infty 2(l+1)
\int_{-\infty}^\infty dx\,e^{ix\delta}{s_l^2(xr/a)+s_{l+1}^2(xr/a)\over
s_l^2(x)+s_{l+1}^2(x)}.
\ee
Here $s_l$ and $e_l$ are modified spherical Bessel functions defined 
in (\ref{eands}) below.
Details of these calculations were given in Ref.~\refcite{cond}.  Graphs of
these condensate densities are shown in Fig.~\ref{fig1}.  The central values 
seem to agree with the observed values of the quark and gluon 
condensates,\cite{vainshtein} but they diverge as the surface of the bag
is approached.

\section{Casimir Forces on Spheres}

The calculations presented here were
carried out in response to the recent program of the  MIT group.\cite{jaffe}
They rediscovered irremovable divergences in the Casimir energy 
for a circle first 
discovered by Sen in 1981,\cite{sen} but then found divergences in the case of
a spherical surface, thereby casting doubt on the validity of the
Boyer calculation for the Casimir energy of a perfectly conducting
spherical shell.\cite{boyer}
  Some of their results, as we shall
see, are spurious, and the rest had been earlier discovered by the 
Leipzig group.\cite{bordag}  However, the MIT group's
work has been valuable in sparking new investigations of the problems of
surface energies and divergences.\cite{fulling}

\subsection{$\delta$-function Potential}
We now carry out a calculation\cite{milton04a} of the zero-point energy 
in three spatial dimensions,
with a radially symmetric singular background
\be
\mathcal{L}_{\rm int}=-\frac12\frac{\lambda}a\delta(r-a)\phi^2(x),
\ee
which would correspond to a Dirichlet shell in the limit $\lambda\to\infty$.
The time-Fourier transformed Green's function satisfies the equation
($\kappa^2=-\omega^2$)
\be
\left[-\nabla^2+\kappa^2+\frac{\lambda}a\delta(r-a)\right]G(\mathbf{r,r'})=
\delta(\mathbf{r-r'}).
\ee
We write $G$ in terms of a reduced Green's function
\be
G(\mathbf{r,r'})=\sum_{lm}g_l(r,r')Y_{lm}(\Omega)Y^*_{lm}(\Omega'),
\ee
where $g_l$ satisfies
\be
\left[-\frac1{r^2}\frac{d}{d r}r^2\frac{d}{d r}+
\frac{l(l+1)}{r^2}+\kappa^2
+\frac{\lambda}a\delta(r-a)\right]g_l(r,r')=\frac1{r^2}\delta(r-r').
\label{redgf}
\ee
The solution inside the sphere ($0<r,r'<a$) is
\be
 g_l(r,r')=\frac1{\kappa r r'}\left[e_l(\kappa r_>)s_l(\kappa r_<)
-\frac{\lambda}{\kappa a}e_l^2(\kappa a)\frac{s_l(\kappa r)s_l(\kappa r')}{1
+\frac{\lambda}{\kappa a}s_l(\kappa a)e_l(\kappa a)}\right].
\label{insphgf}
\ee
Here we have introduced the modified Riccati-Bessel functions,
\be
s_l(x)=\sqrt{\frac{\pi x}2}I_{l+1/2}(x),\quad
e_l(x)=\sqrt{\frac{2 x}\pi}K_{l+1/2}(x).
\label{eands}
\ee
Note that this
reduces to the expected result, vanishing as $r\to a$,
in the limit of strong coupling.
When both points are outside the sphere, $r,r'>a$, we obtain a similar
result, with $e_l\leftrightarrow s_l$ in  the second term.

\newcommand{\bnabla}{\mbox{\boldmath{$\nabla$}}}
\subsection{Pressure on sphere}
Now we want to get the radial-radial component of the stress tensor
to extract the pressure on the sphere, which is obtained by applying the
operator
\be
\partial_r\partial_{r'}-\frac12(-\partial^0\partial^{\prime0}+\bnabla
\cdot\bnabla')\to\frac12
\left[\partial_r\partial_{r'}-\kappa^2-\frac{l(l+1)}{r^2}\right]
\label{radop}
\ee to the Green's function, where in the last term we have 
averaged over the surface of the sphere.
In this way we find, from the discontinuity of
$\langle T_{rr}\rangle$ across the $r=a$ surface, the net stress,
which can alternatively be obtained by differentiating
 with respect to $a$, with $\lambda/a$ fixed, the 
%
total Casimir energy
\be
E=-\frac{1}{2\pi a}\sum_{l=0}^\infty  (2l+1)\int_0^\infty d x\,x\,
\frac{d}{d x}\ln\left[1+\lambda I_\nu(x)K_\nu(x)\right].
\label{teenergy}
\ee
 This expression, upon
integration by parts, coincides with that given recently by 
Barton,\cite{barton}
and was first analyzed in detail by Scandurra.\cite{scandurra}

\subsection{Strong coupling}
\label{sec2.3}
For strong coupling,
the above energy reduces to the well-known expression for the Casimir energy
of a massless scalar field  inside and
outside a sphere upon which Dirichlet boundary conditions are imposed,
that is, that the field must vanish at $r=a$: ($\nu=l+1/2$)
\be
\lim_{\lambda\to\infty}E=-\frac{1}{2\pi a}\sum_{l=0}^\infty (2l+1)\int_0^\infty
d x\,x\,\frac{d}{d x}\ln\left[I_\nu(x)K_\nu(x)\right],\label{dsph}
\ee
because multiplying the argument of the logarithm by a power of $x$ is
without effect, corresponding to a contact term. 
This may be evaluated numerically:\cite{benmil}
$E^{\rm TE}={0.002817}/a$,
which is much smaller than the Boyer result for electrodynamics:\cite{boyer}
$E^{\rm EM}={0.04618}/a$,
although both are repulsive.

\subsection{Weak coupling}
The opposite limit is of interest here.  The expansion of the logarithm
in (\ref{teenergy})
is immediate for small $\lambda$.  The first term, of order $\lambda$, 
is evidently
divergent, but irrelevant, since that may be removed by renormalization
of the tadpole graph.  In contradistinction to the claim of of the MIT 
group\cite{jaffe}
the order $\lambda^2$ term is finite. That term is
\be
E^{(\lambda^2)}=\frac{\lambda^2}{4\pi a}
\sum_{l=0}^\infty(2l+1)\int_0^\infty d x\,x
\frac{d}{d x}[I_{l+1/2}(x)K_{l+1/2}(x)]^2.\label{og}
\ee
The sum on $l$ can be carried out using a trick due to Klich:\cite{klich}
The sum rule
\be
\sum_{l=0}^\infty (2l+1)e_l(x)s_l(y)P_l(\cos\theta)=\frac{xy}\rho e^{-\rho},
\ee
where $\rho=\sqrt{x^2+y^2-2xy\cos\theta}$, 
is squared, and then integrated over $\theta$.
In this way we learn that
\be
\sum_{l=0}^\infty (2l+1)e_l^2(x)s_l^2(x)=\frac{x^2}2\int_0^{4x}\frac{d w}w
e^{-w}.
\ee
Although this integral is divergent, because we did not integrate
(\ref{teenergy}) by parts, 
that divergence does not contribute:
\be
E^{(\lambda^2)}=\frac{\lambda^2}{4\pi a}\int_0^\infty d x\,
\frac12 x \,\frac{d}{d x}\int_0^{4x}
\frac{d w}w e^{-w}=\frac{\lambda^2}{32\pi a},
\label{4.25}
\ee
which is exactly the result I had found earlier,\cite{milton03}
based on the following formula for a hypersphere in $D$ dimensions:
\be
E^{(\lambda^2)}_D=-\frac{\lambda^2}{\pi a}\frac{\Gamma\left(\frac{D-1}2\right)\Gamma(D-3/2)
\Gamma(1-D/2)}{2^{1+2D}[\Gamma(D/2)]^2}.
\ee
which exhibits poles when $D$ is even, where the Casimir energy is known
to diverge.\cite{benmil}

\subsection{Divergences}
However, before we wax too euphoric, we recognize that the order $\lambda^3$ 
term
appears logarithmically divergent, just as the MIT group claimed.\cite{jaffe}
To study the behavior of the sum for large values of $l$, we can use the
uniform asymptotic expansion (Debye expansion),
\be
\nu\gg1:\quad I_\nu(x)K_\nu(x)\sim\frac{t}{2\nu}\left[1+\frac{A(t)}{\nu^2}
+\frac{B(t)}{\nu^4}+\dots\right].
\label{uae}
\ee
Here $x=\nu z$, and $t=1/\sqrt{1+z^2}$.  The functions $A$ and $B$, etc., are
polynomials in $t$.  We now insert this into the energy expression and expand
not in $\lambda$ but in $\nu$; the leading term is
\be
E^{(\lambda^3)}\sim
\frac{\lambda^3}{24\pi a}\sum_{l=0}^\infty\frac1\nu\int_0^\infty
\frac{d z}{(1+z^2)^{3/2}}=\frac{\lambda^3}{24\pi a}\zeta(1).
\ee
Although the frequency integral is finite, the angular momentum sum is
divergent.  The appearance here of the divergent $\zeta(1)$ seems to
signal an insuperable barrier to extraction of a finite Casimir energy
for finite $\lambda$.  The situation is different in the limit
$\lambda\to\infty$.

This divergence has been known for many years, and was first calculated
explicitly in 1998 by Bordag et al.,\cite{bordag} where the second heat kernel
coefficient gave
\be
E\sim \frac{\lambda^3}{48\pi a}\frac1s,\quad s\to0.
\ee
A possible way of dealing with this divergence was advocated in 
Scandurra.\cite{scandurra} Very recently, Bordag and Vassilevich\cite{bv} have
reanalyzed such problems from the heat kernel approach.  They show that this
$O(\lambda^3)$ divergence corresponds to a surface tension counterterm,
an idea proposed by me\cite{milton80} in 1980  in connection
with the zero-point energy contribution to the bag model.  Such a surface
term corresponds to $\lambda/a$ fixed, which then necessarily implies
a divergence of order $\lambda^3$.  Bordag and Vassilevich\cite{bv}
argue that it is perfectly
appropriate to render this divergence finite by renormalization.

\subsection{Boundary layer energy}
Here we show\cite{milton04a} that the surface energy can be interpreted as the bulk
energy of the boundary layer.  We do this by considering a scalar field
in $1+1+d$ dimensions interacting with the background
$\mathcal{L}_{\rm int}=-\frac\lambda 2\phi^2\sigma$,
where
\be
\sigma(x)=\left\{\begin{array}{cc}
h,&-\frac\delta2<x<\frac\delta 2,\\
0,&\mbox{otherwise},
\end{array}\right.
\ee
with the property that $h\delta=1$.
The reduced Green's function satisfies ($\kappa^2=k_\perp^2-\omega^2$)
\be
\left[-\frac{\partial^2}{\partial x^2}
+\kappa^2+\lambda\sigma(x)\right]g(x,x')=\delta(x-x').
\ee
This may be easily solved in the region of the slab, $-\frac\delta2<x<\frac
\delta2$,  ($\kappa'=\sqrt{\kappa^2+\lambda h}$)
\bea
g(x,x')=\frac1{2\kappa'}\bigg\{e^{-\kappa'|x-x'|}
+\frac1{\hat\Delta}\bigg[
(\kappa^{\prime2}-\kappa^2)\cosh\kappa'(x+x')\nonumber\\
\qquad\mbox{}+(\kappa'-\kappa)^2e^{-\kappa'\delta}\cosh\kappa'(x-x')\bigg]
\bigg\},
\label{slabgf}
\eea
\be
\hat\Delta=2\kappa\kappa'\cosh\kappa'\delta+
(\kappa^2+\kappa^{\prime2})\sinh\kappa'\delta.
\ee

We generalize the considerations of Graham and Olum,\cite{graham}
by considering the stress tensor with
an arbitrary conformal term,
\be
T^{\mu\nu}=\partial^\mu\phi\partial^\nu\phi-\frac12 g^{\mu\nu}(\partial_\lambda
\phi\partial^\lambda\phi+\lambda h\phi^2)-\alpha(\partial^\mu\partial^\nu
-g^{\mu\nu}\partial^2)\phi^2.
\ee
We get the following form for the energy density within the slab,
\bea
 T^{00}&=&\frac{2^{-d-2}\pi^{-(d+1)/2}}
{\Gamma((d+3)/2)}\int_0^\infty \frac{d 
\kappa\,\kappa^d}{\kappa'\hat\Delta}\bigg\{(\kappa^{\prime2}-\kappa^2)
\left[(1-4\alpha)(1+d)\kappa^{\prime2}-\kappa^2\right]\cosh2\kappa'x\nonumber\\
&&\mbox{}-(\kappa'-\kappa)^2e^{-\kappa'\delta}\kappa^2\bigg\}.
\eea
From this we can calculate the behavior of the energy density as the
boundary is approached from the inside:
\be
T^{00}\sim \frac{\Gamma(d+1)\lambda h}{2^{d+4}\pi^{(d+1)/2}\Gamma((d+3)/2)}
\frac{1-4\alpha(d+1)/d}{(\delta-2|x|)^d},\quad |x|\to\delta/2.
\label{slabsurfdiv}
\ee  For $d=2$ for example, this agrees with the result found 
by Graham and Olum for $\alpha=0$:
\be
T^{00}\sim\frac{\lambda h}{96\pi^2}\frac{(1-6\alpha)}{(\delta/2-|x|)^d},
\qquad|x|\to\frac\delta2.
\ee
Note that, as we expect, this surface divergence vanishes for the conformal
stress tensor, where $\alpha=d/4(d+1)$.  (There will be subleading
divergences if $d>2$.)  

We can also calculate the energy density on the other side of the boundary,
from the Green's function for $x,x'<-\delta/2$,
\be
g(x,x')=\frac1{2\kappa}\left[e^{-\kappa|x-x'|}-e^{\kappa(x+x'+\delta)}
(\kappa^{\prime2}-\kappa^2)\frac{\sinh\kappa'\delta}{\hat\Delta}\right],
\ee
and the corresponding energy density is given by
\be
T^{00} =-\frac{d(1-4\alpha (d+1)/d)}{2^{d+2}\pi^{(d+1)/2}\Gamma((d+3)/2)}
\int_0^\infty d \kappa\,\kappa^{d+1}\frac{(\kappa^{\prime2}
-\kappa^2)}{\hat\Delta} e^{2\kappa (x+\delta/2)}\sinh\kappa'\delta,
\ee
which vanishes if the conformal value of $\alpha$ is used.
The divergent term, as $x\to-\delta/2$, is just the negative of that found on
the inside.  This is why, when the total energy is computed by
integrating the energy density, it is  finite for $d<2$, and independent
of $\alpha$. The divergence encountered for $d=2$ may be handled by 
renormalization of the interaction potential.
 In the limit as $h\to\infty$, $h\delta=1$, we recover
the divergent expression for a single interface
\be
\lim_{h\to\infty}E_s=\frac1{2^{d+2}\pi^{(d+1)/2}\Gamma((d+3)/2)}
\int_0^\infty d \kappa \,\kappa^d\frac\lambda{\lambda+2\kappa}.
\ee
Therefore, surface divergences have an illusory character.

\subsection{Dielectric Spheres}
The Casimir self-stress on a uniform dielectric sphere was first worked
out by me in 1979.\cite{milton79}  It was generalized to the case when
both electric permittivity and magnetic permeability are present in 
1997.\cite{miltonng}
The result
for the pressure ($x=\sqrt{\varepsilon\mu}|y|$, $x'=\sqrt{\varepsilon'\mu'}|y|$
where $\varepsilon',\mu'$ are the interior, and $\varepsilon, \mu$ are the
exterior, values of the permittivity and the permeability) is
\bea
&&P=-{1\over2a^4}\int_{-\infty}^\infty{d y\over2\pi}e^{i y\delta}
\sum_{l=1}^\infty{2l+1\over4\pi}
\Bigg\{x{d\over d x}\ln D_l\nonumber\\
&&\mbox{}+2x'[ s_l'(x') e_l'(x')- e_l(x') s_l''(x')]
-2x[ s_l'(x) e_l'(x)- e_l(x) s_l''(x)]\Bigg\}\;,
\label{stress}
\eea
where the ``bulk'' pressure has been subtracted, and
\be
D_l=[s_l(x') e_l'(x)-
 s_l'(x')e_l(x)]^2-\xi^2[s_l(x') e_l'(x)+
s_l'(x') e_l(x)]^2,
\ee
with the parameter $\xi$ being
\be
\xi=\frac{\sqrt{\frac{\varepsilon'}{\epsilon}\frac{\mu}{\mu'}}-1}{
\sqrt{\frac{\epsilon'}{\epsilon}\frac{\mu}{\mu'}}+1},
\ee
and $\delta\to0$ is the temporal regulator.
This result is obtained either by computing the radial-radial component
of the stress tensor, or from the total energy.

In general, this result is divergent.  However, consider
the special case
$\sqrt{\epsilon\mu}=\sqrt{\epsilon'\mu'}$, that is, when the speed of light
is the same in both media.
Then $x=x'$ and the Casimir energy reduces to
\be
E=-\frac{1}{4\pi a}\int_{-\infty}^\infty d y\,
e^{i y\delta}\sum_{l=1}^\infty
(2l+1)x\frac{d}{d x}\ln[1-\xi^2((s_le_l)')^2],
\label{special}
\ee
where
\be
\xi=\frac{\mu-\mu'}{\mu+\mu'}=-\frac{\varepsilon-\varepsilon'}
{\varepsilon+\varepsilon'}.
\label{emu}
\ee
If $\xi=1$ we recover the case of a perfectly conducting spherical
shell, treated above. In fact $E$ is finite for all $\xi$.

Of particular interest is the dilute limit, where\cite{klich}
\be
E\approx\frac{5\xi^2}{32\pi a}=\frac{0.099\,4718\xi^2}{2a}, \quad \xi\ll1.
\label{smallxisphere}
\ee
It is  remarkable that
the value for a spherical conducting shell, for which $\xi=1$, is only 7\% 
lower, which as Klich remarks,\cite{klich}
 is accounted for nearly entirely by the next term in the
small $\xi$ expansion.

There is another dilute limit which is also quite surprising.  For a purely
dielectric sphere ($\mu=1$) the leading term in an expansion in powers
of $\varepsilon-1$ is finite:\cite{dds,bordag}
\be
E=\frac{23}{1536\pi}\frac{(\varepsilon-1)^2}a=(\varepsilon-1)^2\frac{0.004\,767}
{a}.
\label{dilutesph}
\ee
This result coincides with the sum of van der Waals energies of the
material making up the ball as Ng and I showed earlier in 1998.\cite{mng98}
The term of order $(\varepsilon-1)^3$ is divergent.\cite{bordag}
The establishment of this result
was the death knell for the Casimir
energy explanation of sonoluminescence.\cite{sono}

\section{Dielectric Cylinders}
The fundamental difficulty in cylindrical geometries is
that there is in general no decoupling between TE and TM modes.
Progress in understanding has therefore been much slower in this regime.
It was only in 1981 that it was found that the electromagnetic Casimir
energy of a perfectly conducting cylinder was attractive, the energy
per unit length being\cite{deraad}
$\mathcal{E}_{\mathrm{em,cyl}}=-{0.01356}/{a^2}$,
for a circular cylinder of radius $a$.  The corresponding result for a
scalar field satisfying Dirichlet boundary conditions of the cylinder is
repulsive,\cite{nest}
$\mathcal{E}_{\mathrm{D,cyl}}={0.000606}/{a^2}$.
These ideal limits are finite, but, as with the spherical geometry, less
ideal configurations have unremovable divergences.  For example, 
 a cylindrical
$\delta$-shell potential has divergences (in third
order).\cite{scand00}  And it is expected that a dielectric cylinder
will have a divergent Casimir energy, although the coefficient of 
$(\varepsilon-1)^2$ will be finite for a dilute dielectric 
cylinder,\cite{bordag01}
corresponding to a finite van der Waals energy between the molecules
that make up the material.  In fact, a calculation of the renormalized
van der Waals energy for a dilute dielectric cylinder gives zero, as is
the Casimir energy for a cylinder for which the speed of light is
the same inside and out, because $\varepsilon\mu=1$.\cite{milton99}  
A calculation
of the Casimir energy for a dielectric cylinder is in progress with
my graduate student Ines Cavero-Pelaez.

\section{Conclusions}

We began by reviewing ancient results for the contributions of zero-point
energies in hadronic (bag) models.
We saw that in general there are surface divergences in the condensates,
and in the energy densities, for the quark and gluon fields.
We discussed $\delta$-function potentials, which in general give divergent
results in 3rd order, but are finite in strong coupling.
These divergences are largely identified with surface energies, which
can be interpreted as bulk energies when the boundaries are smoothed.
 Precisely analogous phenomena happen for dielectric balls and
cylinders (although there is some remarkable symmetry buried in the latter).
I hope this improved understanding of divergences can lead to improved
hadronic models.
For the current status of Casimir phenomena, see my review
article,\cite{milton04} where complete references can be found.

\section*{Acknowledgments}

This work was supported in part by the US Department of Energy.
I thank H. Fried for inviting me to participate in this fascinating meeting.


\end{document}